# Exploring the trilemma of cost-efficient, equitable and publicly acceptable onshore wind expansion planning

Jann Michael Weinand[1,*], Russell McKenna[2,5], Heidi Heinrichs[3], Michael Roth[4], Detlef Stolten[3], Wolf Fichtner[1]

[1]Chair of Energy Economics, Karlsruhe Institute for Technology, Germany

[2]Chair of Energy Transition, School of Engineering, University of Aberdeen, King's College, United Kingdom

[3]Institute of Energy and Climate Research – Techno-Economic Systems Analysis (IEK-3), Forschungszentrum Jülich, Germany

[4]Department of Landscape Planning, Nürtingen-Geislingen University, Germany

[5]Department of Mechanical and Process Engineering (MAVT), ETH Zurich, Switzerland

*Corresponding author: Jann Michael Weinand, jann.weinand@kit.edu, +49 721 608 44444

## Abstract

Onshore wind development has historically focused on cost-efficiency, which may lead to inequitable turbine distributions and public resistance due to landscape impacts. Using a multi-criteria planning approach, we show how onshore wind capacity targets can be achieved by 2050 in a cost-efficient, equitable and publicly acceptable way. For the case study of Germany, we build on the existing turbine stock and use open data on technically feasible turbine locations and scenicness of landscapes to plan the optimal expansion. The analysis shows that while the trade-off between cost-efficiency and public acceptance is rather weak with about 15% higher costs or scenicness, an equitable distribution has a large impact on these criteria. Although the onshore wind capacity per inhabitant could be distributed about 220% more equitably through the expansion, equity would severely limit planning flexibility by 2050. Our analysis assists stakeholders in resolving the onshore wind expansion trilemma.

As the first legally binding global climate change agreement, the Paris Agreement commits approximately 190 parties to preventing climate change and limiting global warming to below 2°C[1]. Some parties, such as Canada, Japan and the European Union, are aiming for climate neutrality by 2050[2,3]. Reducing greenhouse gases through the diffusion of renewable energy technologies can contribute significantly to achieving these objectives[4]. One of the most important renewable energy sources is wind energy with a share of 2.1% in global primary energy consumption[5], which has increased by about 300% between 2010 and 2019[6] and is the highest among renewable energies after hydro power[5]. The already low cost of wind energy is expected to further decrease significantly by 2050[7,8].

While energy system planning predominantly focuses on costs[9], the decentralized nature of renewable energy technologies requires more criteria to be considered. On the one hand, pure cost considerations overlook other opportunities, such as job creation or economic benefits for local communities[10]. For the global power sector, a large increase in jobs is expected through the deployment of renewables by 2050[11]. To ensure that all regions may benefit, an equitable distribution of renewable energy plants is required. On the other hand, despite general approval for renewable energies, local stakeholders increasingly oppose their construction[12,13], especially if they are not involved in the planning process[14,15]. For onshore wind, one of the main reasons for this opposition is the visual impact on the landscape[16–20]. As a result, onshore wind expansion is not accelerating despite declining costs[21]. This is especially evident in Germany, the country with the third largest onshore wind capacity[22] (around 55 GW in 2019[23]) and the fourth largest share of onshore wind in power generation worldwide[24] (about 26%). After record years in 2014 and 2017 with 4.8 GW and 5.3 GW capacity expansions respectively, only 1.0 GW and 1.4 GW new capacity was added in 2019 and 2020[25]. The rapid spread and development of onshore wind turbines has sparked an increase in local protest movements and lawsuits across the country[26,27]. Along with hurdles for new installations introduced by lawmakers, this raises doubts about whether the government's expansion target of an additional 50 GW by 2050 is feasible[23, 26]. A planning approach that addresses the key target criteria of cost-efficiency, public acceptance and equity[28] could underpin the achievement of this target and accelerate the expansion again.

Since the relative weighting of different target criteria is challenging, explorative analyses are needed to compare different spatial optimizations according to individual sustainability criteria[29]. In the literature on national onshore wind site planning, the focus has been mainly on techno-economic criteria. Quantitative analyses that take into account social criteria such as social or political acceptance[30,31] or interregional equity[32–34] are still relatively scarce. A comprehensive review of relevant studies can be found in the Supplementary Material. None of the previous studies have examined the trade-offs between all three criteria cost-efficiency, public acceptance and equity in onshore wind expansion planning.

The objective of this study is to determine optimal locations for onshore wind turbines in 2050. To ensure the lowest possible opposition while maintaining a cost-efficient and equitable onshore wind expansion, we consider all three dimensions of the onshore wind expansion planning trilemma in a multi-objective planning approach and thereby show the trade-offs between the criteria. The onshore wind expansion in Germany serves as a case study for the approach. Prior to the actual optimization, we also examine the existing turbine population to show the relevance of the target criteria considered here. A key challenge for science and practice lies in quantifying public acceptance for onshore wind projects, whose approval depends to a large extent on the scenicness of surrounding landscapes[24]. Therefore, we employ the scenicness of landscapes as a proxy for the landscape impact, itself a significant part of public acceptance of wind installations. Since electricity networks have an impact on both cost-efficiency[35] and landscape (and thus public acceptance[12]), the length of the necessary additional network is also measured as the distance to the nearest transformer. In addition to the scenario with a capacity expansion by 50 GW (German government's 2050 target), we consider an ambitious scenario with an expansion by 145 GW, in line with the call of the German Wind Association for an annual addition of at least 4.7 GW to meet climate targets[36].

**Costs and scenicness define turbine locations**

The share of already existing turbines as a fraction of the technical German onshore wind potential generally decreases as levelized cost of electricity (LCOEs), scenicness of landscapes, or network length increase (Figure 1), which emphasizes the relevance of these planning criteria. The technical potential corresponds to the wind power generated within an available area for wind turbines. It considers constraints such as wind turbine characteristics, wind farm array losses and electrical conversion losses[37]. Figure 1 further shows that much of this potential is still available at favorable locations. The technical potential in Germany in 2050 includes approximately 160,000 onshore wind turbines[38]. About 29,000 turbines have already been installed and are in operation today (about 55 GW in 2019[23], Figure 8 in the method section).

The intensified consideration of the LCOEs, scenicness and networks has led to an uneven distribution of existing turbines across Germany. The turbines are mainly located in the northern federal states Lower Saxony (22%, 15.1 turbines per 1,000 km², Figure 2), Brandenburg (14%, 14.3 turbines per 1,000 km²), North Rhine-Westphalia (13%, 14.0 turbines per 100 km²) and Schleswig Holstein (11%, 22.7 turbines per 1,000 km²), which show the highest capacity factors in Germany[39]. The associated lower cost of wind electricity supply is one reason for the high wind diffusion in the north. Furthermore, the scenicness is lower there than in the southern federal states like Bavaria and Baden-Württemberg (Figure 2), which could indicate a lower resistance towards onshore wind[40]. Except for the sea coast, the lake districts in north-eastern Germany and some local "hot-spots" like the Lueneburg Heath, northern Germany tends to be an area of low to medium scenicness[41].

The existing wind turbines in Germany are located at sites with a mean scenicness of 4.25, which is below the German average of 4.5 (Figure 8 in the method section). Most are located at sites with a scenicness of 3 (15%), 4 (57%) or 5 (16%) (Figure 1). At sites with a scenicness of 1 and 2, as well as 8 and 9, the smallest proportion of turbines is located (<1%). On the one hand this could be due to the exclusion of these areas (cities) because of minimum distance restrictions, and on the other hand because of the low frequency and the high beauty of these areas. At the same time, the highest average share of existing turbines shows that the potential in areas with a scenicness of 1 and 2 has been exploited the most so far (Figure 1).

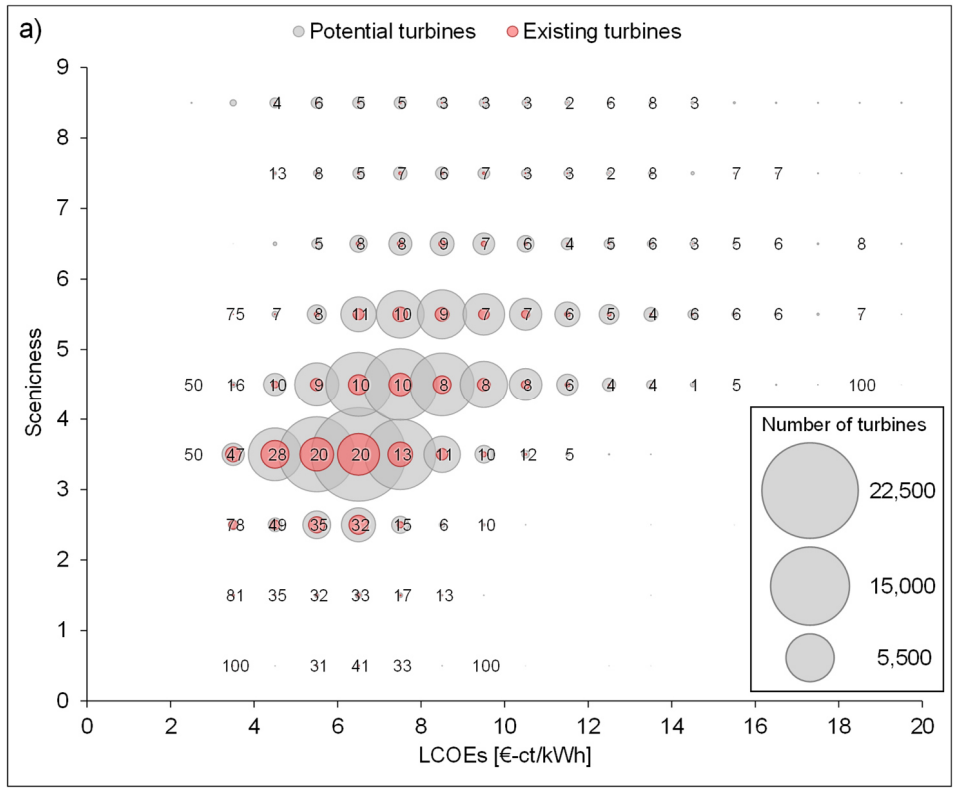

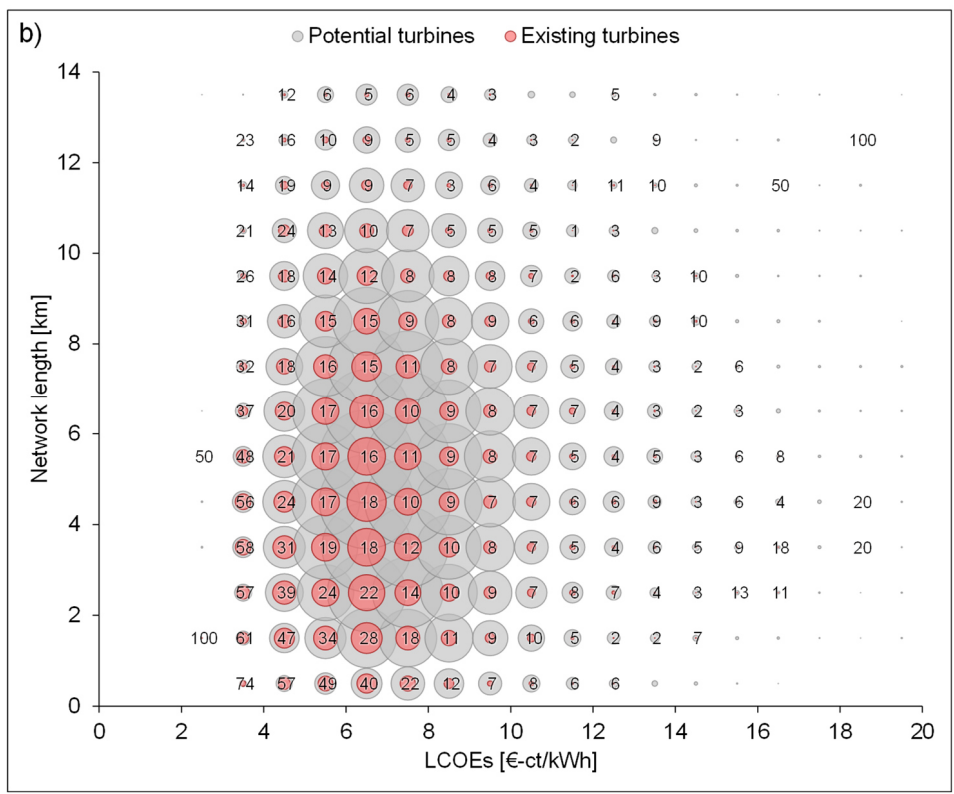

*Figure 1: Number of turbines that could potentially be placed in Germany in 2050 (gray bubbles) and the associated ranges of LCOEs (a and b), scenicness (a) and network length (b). The red bubbles show the number of turbines already installed in Germany and the numbers in the center of the bubbles show their share in the potential, i.e. 100 means that the potential has already been fully exploited. Each bubble applies to an interval, i.e. a bubble between the values 1 and 2 shows the number of turbines at the range (1;2].*

Probably due to lower capacity factors and higher scenicness, the onshore wind diffusion is lowest in the southern federal states Bavaria (1.9 turbines per 1,000 km², Figure 2) and Baden-Württemberg (2.6 turbines per 1000 km²). In general, the mean scenicness in all federal states seems to correlate with the existing specific capacity per km² (Figure 2, correlation coefficient: -0.66). At the same time, the share of lawsuits against wind turbines[42] (measured in terms of total approved capacity) is higher in the southern federal states than in the northern ones (Figure 2). The high differences in the specific number of turbines indicate that there has not yet been an equitable distribution of wind turbines in Germany. In fact, our analysis shows that the regional equity of existing wind capacity per inhabitant measured with the Gini index[43] has a value of only 6.4%, with 100% being a completely equitable distribution.

The remaining technical onshore wind potential in Germany, which could be added to the existing capacity, shows the lowest LCOEs in landscapes with a scenicness of 3 (up to a capacity of about 15 GW), 4 or 5 (Supplementary Figure S9). Since the existing turbines are mainly located in these landscapes, it is evident that cost-efficiency has been the most important factor in siting so far. The greatest onshore wind potential can be realized in landscapes with a scenicness of 4, 5 or 6. In some landscapes with scenicness below average, significantly higher LCOEs have to be expected.

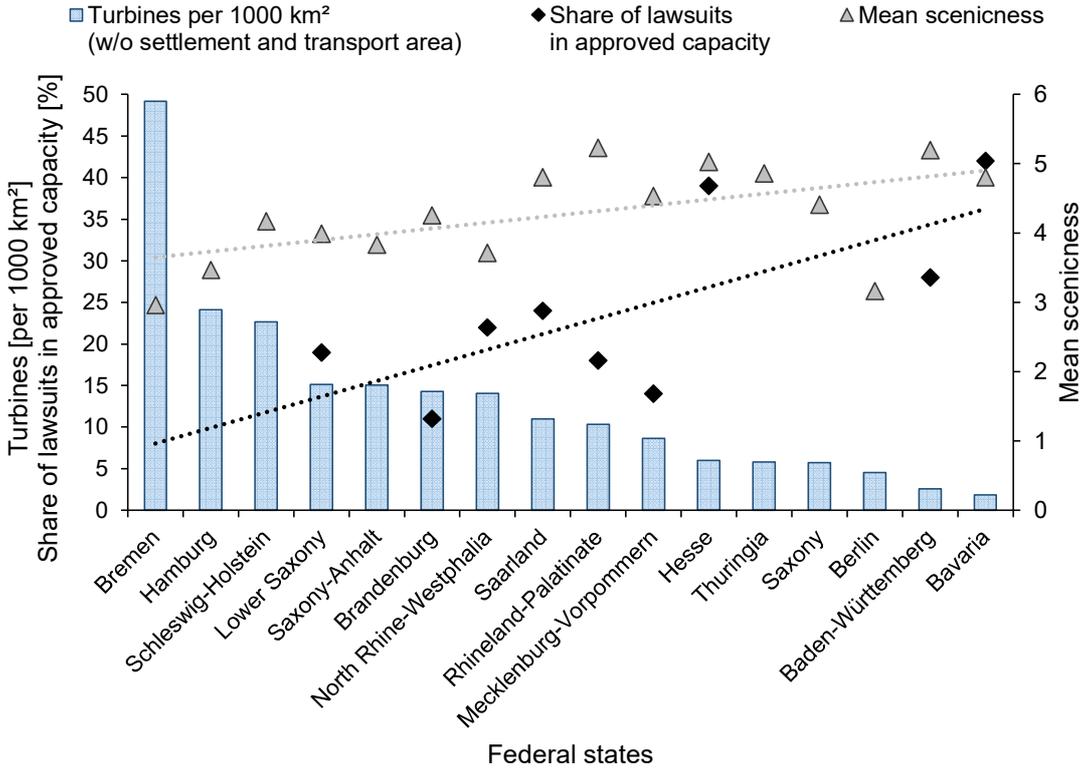

*Figure 2: Specific number of turbines, share of wind turbines in total licensed capacity subject to lawsuits, and mean scenicness of landscapes in the sixteen German federal states. The federal states are ordered from the highest specific number of turbines (Bremen) to the lowest (Bavaria). The share of wind turbines in total licensed capacity subject to lawsuits has been determined by a survey with 89 companies in 14 of the sixteen federal states[42]. However, lawsuits have not been reported for all federal states[42].*

**Weak trade-off between cost-efficiency and public acceptance**

The trade-offs that emerge between optimal cost-efficiency and public acceptance are rather weak. The expansion of onshore wind turbines with a total capacity of about 50 GW would be associated with mean LCOEs of at least 4.7 €-ct/kWh (scenario Base_LCOE, Table 1), a mean scenicness of at least 3.6 (scenario Base_Scenic) or a mean length of power cables of at least 1.5 km (scenario Base_Network). If the turbines are placed in areas with minimal scenicness, the average LCOEs would only increase by 13%, and in the case of optimal cost-efficiency, the average scenicness would only increase by 17%. As can be seen in Figure 3, the turbine locations in the two cases a) and b) do not differ much either, with around 60% of the locations being "no-regret" sites. "No-regret" site means, that a turbine at a specific location is installed under both optimization criteria[29]. However, in the cost-optimal case, for example, many turbines would be placed in the south in the less steep areas near the Alps, which have very high scenicness. This would change in the case with minimal scenicness; here, more turbines would be placed in central Germany instead.

Table 1: Overview of the 14 scenarios considered and their results for mean LCOES, mean scenicness and mean network length as well as equity. The scenarios are distinguished by their target criteria (minimization criteria), the capacity target for onshore wind in 2050 and whether the expansion has to be equitable or not.

| Scenario | Minimization criteria | Equity included? | Onshore wind capacity [$GW_{2050}$] | Mean LCOEs [€-ct/kWh] | Mean Scenicness | Mean network length [km²] | Equity [%] |
|---|---|---|---|---|---|---|---|
| Base_LCOE | LCOEs | × | 105 | 4.7 | 4.2 | 6.1 | 7.5 |
| Base_Scenic | Scenicness | × | 105 | 5.3 | 3.6 | 5.4 | 7.1 |
| Base_Network | Network length | × | 105 | 7.7 | 4.7 | 1.5 | 9.2 |
| Base_all | All criteria | × | 105 | 5.5 | 3.8 | 3.1 | 7.3 |
| Base_LCOE_E | LCOEs | yes | 105 | 7.0 | 5.2 | 5.6 | 20.5 |
| Base_Scenic_E | Scenicness | yes | 105 | 7.3 | 4.9 | 5.5 | 20.5 |
| Base_Network_E | Network length | yes | 105 | 7.8 | 5.2 | 4.6 | 20.5 |
| Base_all_E | All criteria | yes | 105 | 7.2 | 5.1 | 4.8 | 20.5 |
| High_LCOE | LCOEs | × | 200 | 5.4 | 4.3 | 6.1 | 10.2 |
| High_Scenic | Scenicness | × | 200 | 5.9 | 3.9 | 5.9 | 8.5 |
| High_Network | Network length | × | 200 | 7.5 | 4.8 | 2.6 | 11.5 |
| High_LCOE_E | LCOEs | yes | 200 | 6.3 | 4.8 | 5.8 | 16.1 |
| High_Scenic_E | Scenicness | yes | 200 | 6.8 | 4.5 | 5.6 | 16.2 |
| High_Network_E | Network length | yes | 200 | 7.6 | 4.9 | 3.7 | 18.1 |

The required mean length of the electricity network to connect the wind turbines with transformers is relatively high in scenarios Base_LCOE and Base_Scenic with 6.1 km and 5.2 km, respectively, which would affect both the cost-efficiency and the scenery of the landscape. If the distance to the nearest transformer is minimized (scenario Base_Network), the mean LCOEs of the turbines (+64%) and the scenicness of the locations (+31%) change significantly. However, the distribution of turbines in this case would be more equitable: while in the first two cases turbine expansion still occurs mainly in the north of Germany, in the case with minimum distance to the transformers we now see a much more even expansion across the German territory (Figure 3c). Compared to the Base_LCOE and Base_Scenic scenarios, only about 6% of the turbines would be installed at "no-regret" sites. At the same time, the mean network length decreases by up to 75% (Table 1).

These previously described trade-offs would behave similarly even with a higher target capacity of 200 GW instead of 105 GW in 2050 (scenarios High_LCOE, High_Scenic and High_Network, Table 1). However, the minimum mean LCOEs, scenicness and network lengths would increase by 15%, 8% and 7%, respectively. Furthermore, in High_LCOE and High_Scenic, significantly more turbines would no longer be located only in the north of Germany as in Base_LCOE and Base_Scenic, which would increase equity (Supplementary Figure S10a-b).

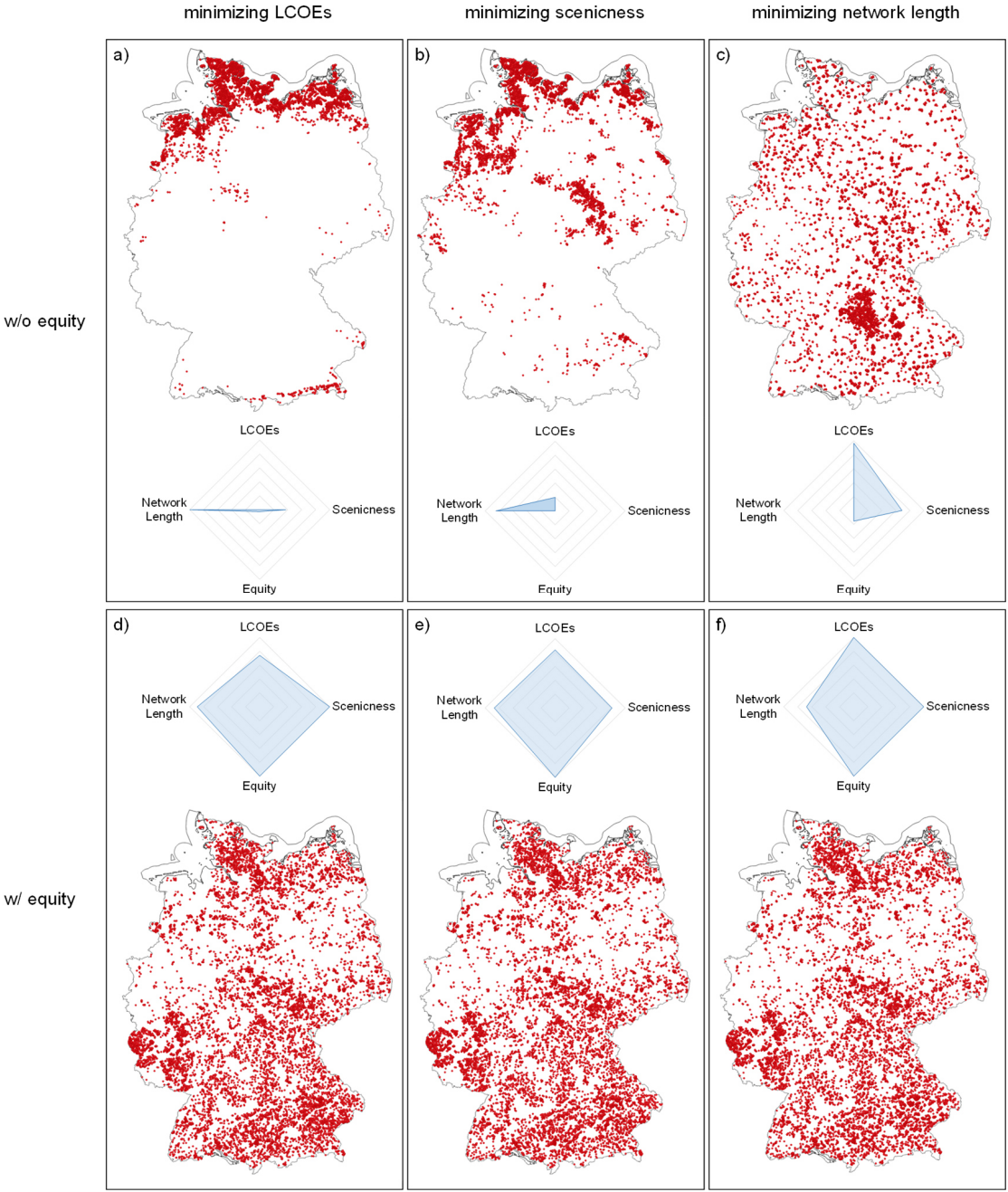

*Figure 3: Optimal locations of onshore wind turbines to be added by 2050 for different target criteria. Turbines are shown as red dots. Around 50 GW of capacity is added in each case. a) – c) show the optimal locations when minimizing LCOEs (scenario Base_LCOE), scenicness (Base_Scenic) or network length (Base_Network), respectively. In d) – f) the same target criteria are used, but in these cases with the constraint, that the capacity expansion has to be equitable (Base_LCOE_E, Base_Scenic_E and Base_Network_E). The values in the spider charts are scaled based on the minimum and maximum values among the "Base" scenarios.*

**Inconsistent relationships between windy and scenic locations**

The weak trade-off between cost-efficiency and scenicness found in Germany is not observed in other regions. This is particularly interesting when compared with a recent study, which economically assesses the technical onshore wind potential of Great Britain (GB) as a function of scenicness[35]. Whilst for Germany, the scenicness data covers for the entire land area[41], for GB the data exist on a 1 km squared grid distributed throughout the country[44]. This allows comparison to the technical potential data we use for Germany in the present analysis (Figure 4). The common feature of both GB and Germany is the distribution of the largest potentials among the mean scenicness values 4, 5 and 6. However, in contrast to Germany, the LCOEs of wind turbines decrease almost linearly as a function of scenicness in GB. While in Germany the beautiful landscapes in the south – except for the very beautiful ones with scenicness 9, e.g. in the foothills of the Alps – have higher LCOEs than the windy but less beautiful north, in GB the highest scenicness is mainly in the north of Scotland with beautiful landscapes and high capacity factors[35]. In the German context, this relationship is favourable, because it implies a complementarity (rather than competition) between windy and scenic locations.

However, we demonstrate below that in some German regions (e.g. Bavaria) the trade-off between cost-efficiency and scenic locations is indeed rather strong. Also, the top six locations with the best wind resources and thus lowest LCOEs in Germany are found at a scenicness value of 9. Furthermore, among all scenicness categories, the share of LCOEs smaller than 5 €-cent/kWh is highest for the scenicness category 9 (17%).

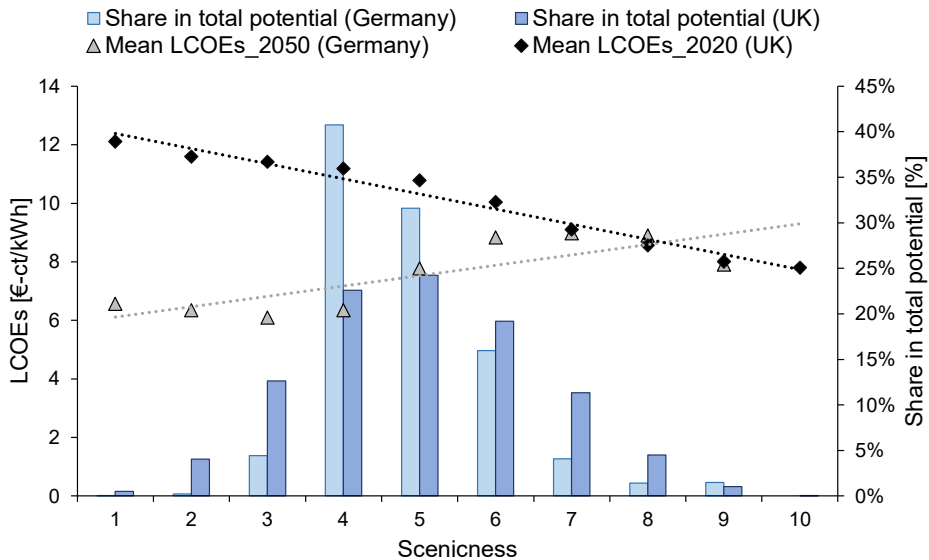

*Figure 4: Mean LCOEs and shares in total onshore wind potential for different scenicness classes of landscapes in Great Britain and Germany. For the scenicness dataset of Great Britain, the ratings range from 1 (low) to 10 (high), for Germany from 1 (low) to 9 (high). In addition, the trend lines of the mean LCOEs as a function of scenicness are shown (black: y=-0.516x+12.902, R²=0.97; grey: y=0.355+5.755, R²=0.65).*

**Equity significantly constrains planning options**

The Pareto curves in Figure 5 relate two of the target criteria to each other, and show only a rather weak trade-off between LCOEs and scenicness in the case without equity. If the network length is included, as already described, there is a large influence on LCOEs or scenicness. It is interesting to note that the network length can be reduced significantly without a considerable increase in the mean scenicness. In general, these curves illustrate that small losses in one target criterion can result in significant improvements in another.

In the case of equitable expansion planning, these effects do not appear. While the slopes of the Pareto curves in Figure 5 are similar to the cases without equity, the range in the target criteria values is now significantly smaller: the mean LCOEs can now only change by up to 11% instead of 64%, the mean scenicness by up to 6% instead of 31%, and the mean network length by up to 22% instead of 306%. Hence including equity leads to a significantly smaller planning flexibility. The locations of the turbines would therefore be practically fixed as the comparison of Figure 3d-f further demonstrates. Whilst mean LCOEs, scenicness and network length increase significantly in the scenarios Base_LCOE_E, Base_Scenic_E and Base_Network_E in comparison to the scenarios without equity by up to about 50%, 35% or 205%, respectively, the turbines are now distributed much more equitably: Compared to the current distribution of the existing turbine stock, equity increases by about 220%. However, the equity reaches a maximum value of only up to 20.5% as due to the current uneven distribution of existing turbines and low or lacking potentials in many regions, an equity value of 100% is far from achievable with only onshore wind. This is further demonstrated by the scenarios High_LCOE_E, High_Scenic_E and High_network_E, which show lower equity values between 16% and 18% despite almost twice as much capacity. In these scenarios the potential maximum capacity is reached in many regions and therefore more capacity has to be installed in regions that already have a high capacity. However, due to the increased capacity, the planning flexibility is higher (Supplementary Figure S10d-f), and mean LCOEs, scenicness, and network length can change by as much as 21%, 9%, and 57%, respectively, at different target weights.

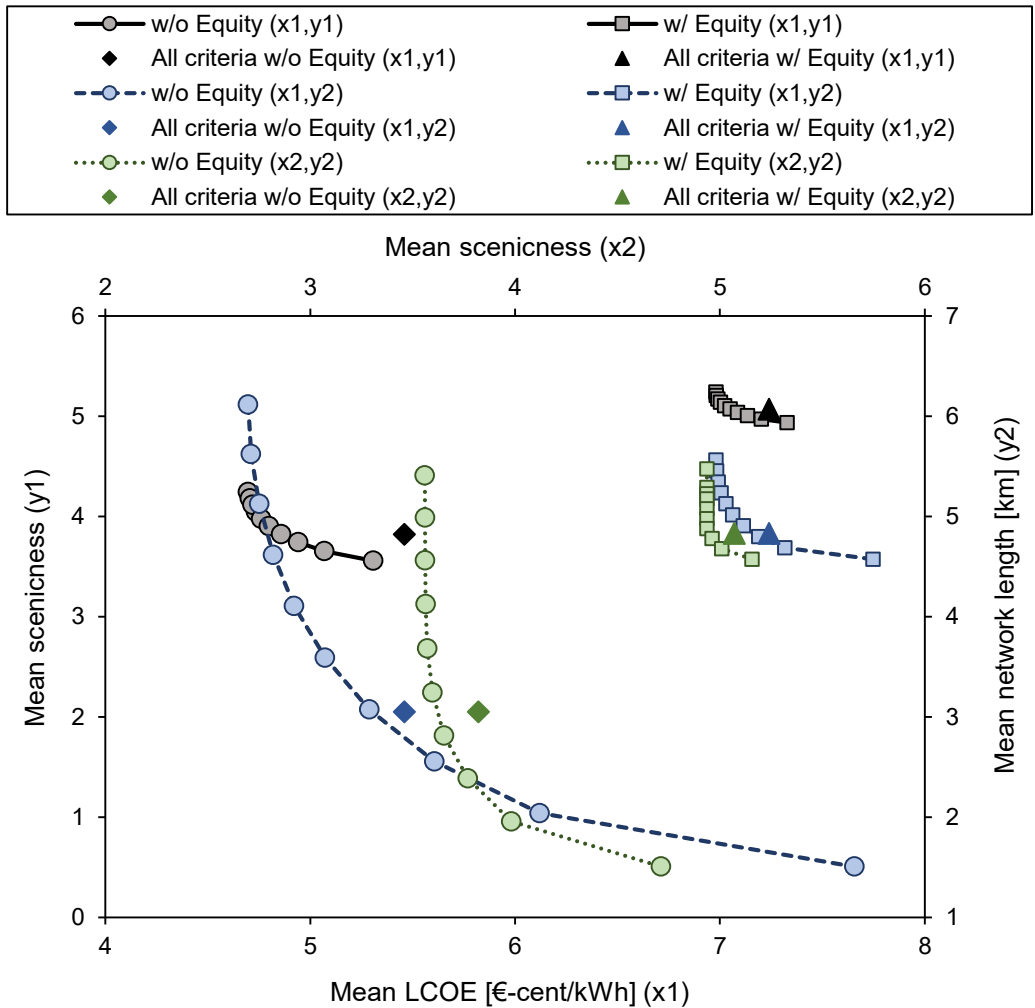

*Figure 5: Pareto fronts between different target criteria. The curves and points each relate to two axes indicated in the legend. The grey curves represent the pareto fronts between scenario Base_LCOE and Base_Scenic, and between Base_LCOE_E and Base_Scenic_E, the blue curves between Base_LCOE and Base_Network, and between Base_LCOE_E and Base_Network_E, and the green curves between Base_Scenic and Base_Network, and between Base_Scenic_E and Base_Network_E. The minimum target values of the criterion to be achieved on the y-axis were decreased by 10% in each optimization, and then the minimum value of the criterion on the x-axis was determined. The diamonds and triangles show the optimum with simultaneous minimization of all target criteria in the case without and with equity, respectively.*

When all criteria are simultaneously optimized with equal weightings rather than just one criterion, the locations of scenarios Base_all and Base_all_E in Figure 6 result. Base_all includes turbines from all three scenarios Base_LCOE, Base_Scenic and Base_Network: some cost-efficient turbines in landscapes with low scenicness in the north; further turbines in central Germany, which were chosen mainly when minimizing scenicness; and a few turbines in southern Germany, which were chosen when minimizing the required network length. Except for the Pareto curve for scenario Base_LCOE and Base_Scenic, the optimum of Base_all lies pretty much in the middle of all Pareto curves (Figure 5). Compared to the minima in scenarios Base_LCOE, Base_Scenic and Base_Network, the mean LCOEs would increase by 17%, the mean scenicness by 6%, and the mean network length by 107%. Therefore, when minimzing all criteria, the trade-offs between the objective criteria are lower except for cost-efficiency, which increases less from scenario Base_LCOE to Base_Scenic (+13% instead of +17%).

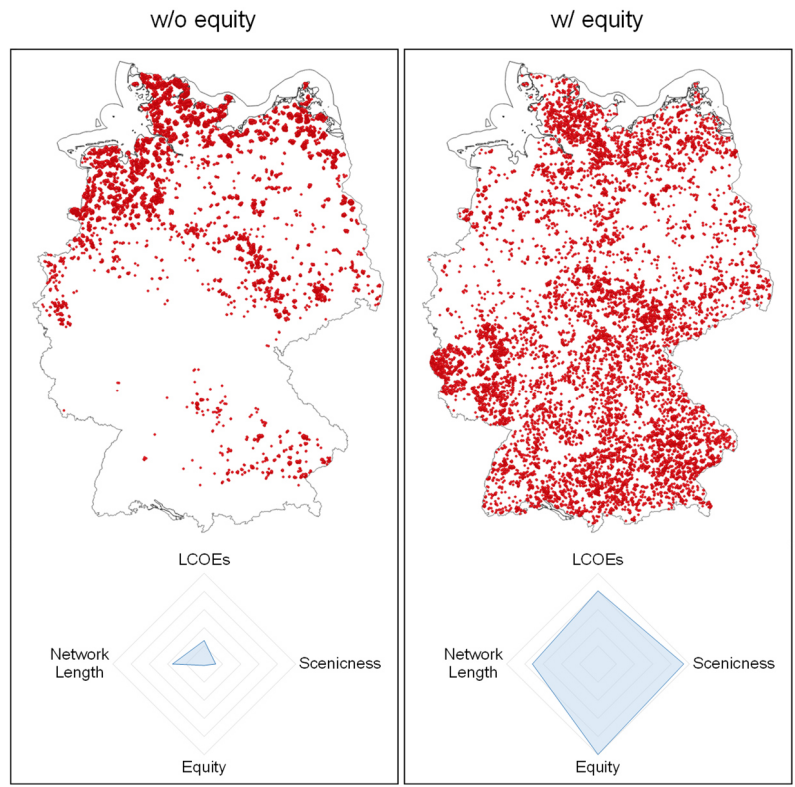

*Figure 6: Optimal locations of onshore wind turbines to be added by 2050 if all three target criteria are considered. Turbines are shown as red dots. Around 50 GW of capacity is added in each case. The left part shows the optimal locations without considering equity (Base_all), the right part with considering equity (Base_all_E). The values in the spider charts are scaled based on the minimum and maximum values among the "Base" scenarios.*

As already seen in Base_LCOE, Base_Scenic and Base_Network, also in Base_all the mean objective values increase significantly if an equitable distribution should be achieved. When comparing the LCOEs, scenicness and network length at the locations where the new turbines are placed in Base_all and Base_all_E (Supplementary Figure S11), it is obvious that the most advantageous locations are no longer exploited in Base_all_E. However, the trade-offs in Base_all_E are smaller: compared to the minimum values in Base_LCOE_E, Base_Scenic_E and Base_Network_E, the mean LCOEs, scenicness and network length increase only by 3%, 4% and 4%, respectively. However, this is also related to the low planning flexibility in the scenarios with equity.

**Subordinate political targets only achievable with equitable expansion**

While the policy target for onshore wind capacity in 2050 can be met in any scenario, subordinate targets such as the "south quota" can only be achieved in scenarios that involve a more equitable distribution of turbines. In recent years, the northern-focused expansion of onshore wind resulted in high and increasing amounts of curtailed electricity, with more than 5 TWh in 2019[45]. Curtailment means the deliberate reduction of output power below the level that could be generated to balance energy supply and demand or due to transmission constraints[46]. In the Renewable Energy Sources Act 2021, a minimum south quota of 15-20% of new wind development over 2021-2024 will be established to address this issue[47]. Currently, the south quota is below 10% and would further reduce in most scenarios without equity due to the fact that, as has been shown, the cost-optimal turbine locations would still be in the less beautiful landscapes in the north of Germany (Figure 7). However, in the scenario Base_Network, as well as all scenarios with an equitable expansion, the south quota could be increased to a value of about 40%. Therefore, apart from higher costs and lower planning flexibility, the equity scenarios would mostly avoid further curtailment.

As Figure 7 demonstrates, onshore wind expansion in the equity scenarios would be largely in the states of Baden-Württemberg and Bavaria, where the fewest wind turbines relative to the area have been built to date (Figure 2) and which face particularly strong opposition to onshore wind[40]. We have shown above a generally weak trade-off between cost-efficiency and scenicness in Germany, and thus a rather low opposition to onshore wind should be expected in a national planning context. However, the weak trade-off does not apply to Bavaria and Baden-Württemberg: the mean scenicness in these states deviates by 66% with 8.7 (Base_LCOE) and 2.9 (Base_Scenic). In the two scenarios with equity, the difference would be only 6% with 5.6 (Base_LCOE_E) and 5.2 (Base_Scenic_E). All in all, it could be feasible to achieve the south quota – especially in the scenario Base_Scenic_E, in which the turbines are placed in less beautiful landscapes in Bavaria and Baden-Württemberg.

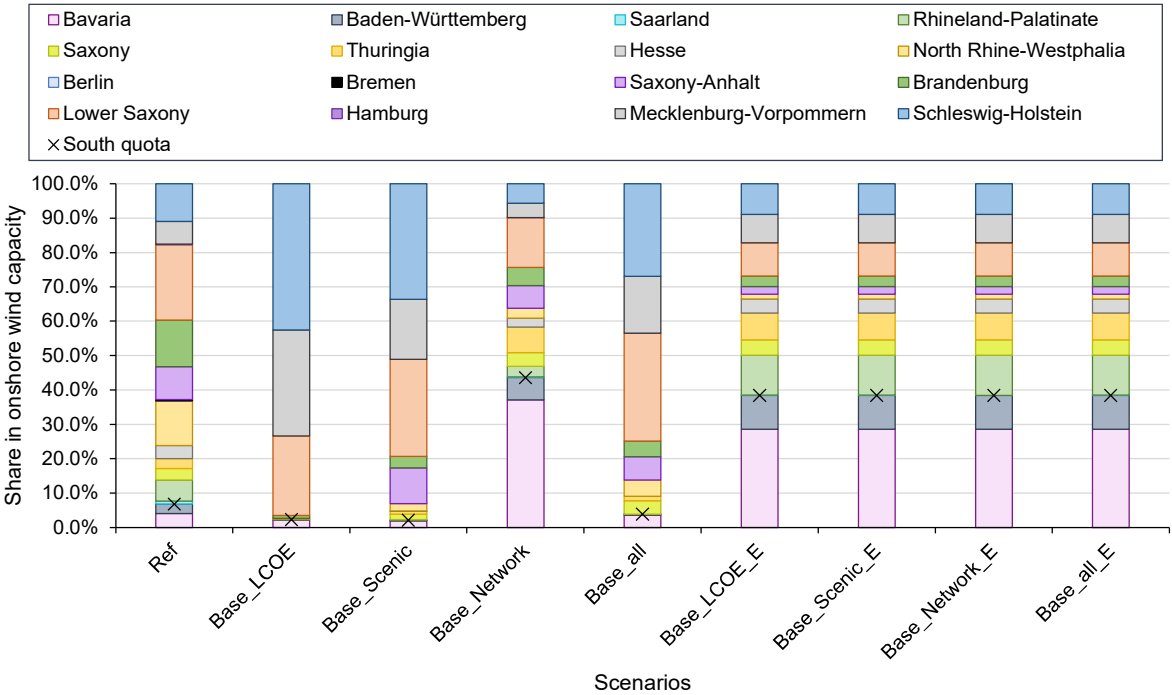

*Figure 7: Shares of the sixteen federal states in onshore wind capacity in 2020 and in eight scenarios for 2050. "Ref" shows the shares of existing capacity, the other scenarios show only the shares of added capacity by 2050. The federal states are sorted from south (bottom) to north (top).*

**Discussion**

The targeted onshore wind expansion to achieve the 2050 climate targets will face a number of obstacles. The three most relevant criteria for the successful diffusion of turbines form the trilemma in onshore wind expansion planning: cost-efficiency, public acceptance and equity. We combined a variety of open data sets, a Geographic Information System, and a multi-criteria optimization to determine the trade-offs between these three objectives within an optimal onshore wind expansion plan. As a necessary condition for the implementation of the expansion in reality, public acceptance was quantified for the first time in an expansion planning framework. As a proxy for public acceptance, the scenicness of landscapes was used, since the impact on landscapes is the most prominent motivation for the opposition towards onshore wind.

In our analysis, a number of assumptions may partly affect the results. We considered only onshore wind, which means that we neglected the opportunity costs of the required land. In addition, other renewable technologies like solar energy, which could be more affordable in the southern regions of Germany, were neglected. However, the German government has specific plans for onshore wind expansion, also regarding its geographic locations, especially regarding southern Germany, as demonstrated above.

Furthermore, our analysis is static. First, this means we ignored the deployment process for the turbines. Second, we only considered capacity and not hourly generation and its volatility, since the onshore wind targets are formulated as capacities by the government. Third, repowering is expected to become increasingly important for the wind industry in the future[48]. We neglected repowering of existing wind farms, despite turbines probably having a capacity of more than 5 MW in 2050 instead of the current mean of about 2.5 MW[8]. Since the higher performance turbines would also require greater minimum distances and thus more land area, this effect was evaluated as remaining constant.

We considered only the straight line distance for connecting the wind turbines with transformers, which is common practice in electricity network planning. Often, a detour factor is used, to account for locations, in which the straight line route is not feasible[49].

In addition, we define an equitable distribution of wind turbines by using population size. However, criteria could also include the ratio between wind energy potential and actual generation[32], direct employment for electricity generation and storage[34], or direct land use per total area[34]. In another article on the spatial allocation of renewable power plants[32], equity is also weighted using population size. Since public acceptance is relevant in the present study, a distribution of capacity based on population size is assumed to be equitable. If other criteria such as average income are used, then poorer regions that do not aspire to have wind turbines in their vicinity might be disadvantaged. The population size is therefore chosen as the most suitable criterion from an objective point of view.

In the following, the key findings of the analysis for the German case study are discussed. **First**, our analysis of the existing turbine stock shows that the criteria used in our planning tool are of high relevance in reality. With increasing LCOEs, scenic beauty of the landscape or required network length, the share of existing turbines in the onshore wind potential (at the respective locations) decreases. However, this also means that an equitable expansion has been mostly neglected so far, see below.

**Second**, the trade-off between cost-efficiency and beauty of the landscapes in which the turbines would be placed, and thus public acceptance, turns out to be rather weak in Germany. Considering these target criteria, in both cases the turbines would be installed mainly in the northern federal states Lower Saxony, Mecklenburg-Vorpommern and Schleswig-Holstein (Base_LCOE and Base_Scenic in Figure 7). These are the same states that already account for the largest share of existing capacity (Ref in Figure 7), meaning that cost-efficiency and public acceptance were apparently priorities for historical wind farm developments. In other words, taking these target criteria into account alone would reinforce the unequal onshore wind distribution between North and South.

**Third**, whilst Germany shows a complementarity (rather than competition) between windy and scenic locations, another study[35] for Great Britain showed just the opposite. But there are also individual regions in Germany, such as the federal states Bavaria and Baden-Wuerttemberg, which very well show a higher trade-off between cost-efficiency and scenicness in potential wind sites. On the one hand, this shows that general conclusions cannot be drawn from analyses on the relationship between LCOEs and scenicness for one country. Instead, quantitative bases for scenicness in other countries must also be determined. On the other hand, this assessment of the trade-off between cost efficiency and scenicness of onshore wind strongly depends on the considered system boundaries.

Although the quality of the scenicness assessment model is high and a validation approach with external data has confirmed the validity of the scenicness dataset[50], other factors have an influence on perceived scenicness and thus on the resistance towards wind turbine deployment. Subjective feelings and preferences, as well as place-attachment and local identities also play an important role for landscape appreciation, but could not be incorporated into the scenicness model. Other models for the German-wide assessment of scenic attractiveness give a higher relative weight to water features[51], which leads to higher scenicness values in the north-German plains. As the distribution of optimal locations for wind turbines is sensitive to a modified spatial distribution of scenicness values, different scenicness datasets could lead to a shift of optimal location for wind turbines towards the south.

Furthermore, landscapes with similar scenicness values may have different sensitivities to impacts caused by wind turbines. Thus, other factors such as intervisibility and visual openness should also be considered as proxies for visual landscape sensitivity to wind energy[52,53]. In addition to perceived scenic quality of landscapes, social acceptance of wind turbines is also affected by the recreation potential of landscapes[54]. Finally, while wind turbines and transmission grid infrastructure have a negative influence on perceived scenicness[41], younger generations hardly consider wind turbines to be a general landscape annoyance[55]. This is not true – at least not to the same degree – for transmission lines[56,57]. Thus, the interrelation of network length and scenic landscape quality requires a careful weighing in for future research.

Whilst the landscape impact is arguably most important for public acceptance of onshore wind[16–19], public concern is reduced when the affected individuals live farther from turbines[18,58], or have prior experience with wind energy[59–62]. The quantification of these aspects would also be pertinent to future energy system analyses.

**Fourth**, an equitable expansion is associated with significantly higher costs and higher scenicness at the wind turbine sites as well as a low planning flexibility. The question that arises from our analysis is whether benefits such as regional economic stimulation can outweigh the higher costs, presumably greater public opposition, and lower planning flexibility.

As our analysis further shows, an expansion of onshore wind could only achieve a maximum equity of about 20%. This is partly due to the fact that we apply a brown field approach and take into account the existing turbine stock, which shows an equity of only about 6%. Furthermore, in many German regions there is no or only very limited onshore wind potential due to minimum distance restrictions or technical constraints. However, other technologies could also measure equity: for example, an equitable and cost-efficient distribution may involve onshore wind turbines in the north and photovoltaic panels in the south (mainly in Bavaria and Baden-Württemberg)[32]. But considering that solar photovoltaics has a lower impact on landscapes[63] and leads to less public opposition than onshore wind[64–66], this capacity-based equity with various technologies is not necessarily socially equitable. Even in the north of Germany, where the onshore wind turbines are currently mainly located, local citizens may be concerned that wind turbines might put off tourists and thus negatively affect local incomes[26]. Whilst multi-technology approaches are valuable to consider the interactions of different technologies and criteria, the exclusive focus on onshore wind here is the strength of this study, which adds to the discussion about onshore wind development.

**Fifth**, some subordinate targets, such as the south quota of wind turbines in Germany, can only be met in the expansion scenarios with a more equitable distribution of turbines. This target is necessary to reduce curtailment and transmission grid expansion necessitated by overcapacity in the north. Historically, the diffusion of wind turbines in the south of Germany has been slowed down by local opposition[40]. Here, in particular, the scenarios we have shown for placing wind turbines in less beautiful landscapes could reduce opposition.

**Sixth**, small reductions in one target criterion could result in significant improvements in another. For example, the mean network length required to connect the turbines to transformers can be greatly reduced if the mean scenicness is slightly increased. In general, however, when minimizing turbine LCOEs or scenicness, the mean network length is high. Previous analyses have shown that the networks have a strong influence on total LCOEs (which would double on average if network costs are included[35]) and also on the landscape scenery and thus public acceptance[12]. In this study, the impact of network cables on LCOEs and scenicness was not quantified, in order to determine the optimal turbine locations with a limited number of assumptions. When considering network costs in the LCOEs, wind turbines would have to be clustered into wind farms with heuristics[35], requiring only one connection to the transformers. In the case of scenicness, the impact of the cables on the landscape scenery compared to the turbines would have to be weighted first. Both were indirectly considered with our approach in simultaneously minimizing all three target criteria LCOEs, scenicness, and network length. However, an improvement to the equally weighted consideration of the target criteria through expert elicitation weights faces high hurdles: whilst stakeholders consider interregional equity an important criterion for allocation, agreement on uniform weightings of various criteria for onshore wind expansion by experts appears to be practically impossible[29].

**Lastly**, the rapid spread and development of onshore wind in the past has sparked an increase in local protest movements and lawsuits across the country[26,27]. Our approach, which includes scenicness of landscapes and equity, can assist in moderating such protests. In addition to this, the introduction of the new "Investment Acceleration Act", which ensures that pending lawsuits will no longer halt the planning or construction of onshore wind farms[67], could also accelerate the German onshore wind expansion.

Furthermore, policy makers could consider more ambitious targets for onshore wind expansion as in our "High" scenario with a capacity of 200 GW by 2050. In the past, long-term onshore wind capacity targets have been repeatedly increased in response to developments in the energy sector. In addition, many experts see the achievement of climate protection targets at risk if the current targets of the Renewable Energies Act 2021 are maintained[36]. A recent study has shown that an early and steady decarbonisation of the European energy system would be more cost-effective than a late and rapid path[68]. Also for onshore wind, an early commitment to higher capacities would enhance planning security, prevent conflicts that may arise in the future, and create more planning flexibility as shown by our analysis – even in scenarios with an equitable approach.

**Methods**

In a study of the Federal Ministry for Economic Affairs and Energy[69] the onshore wind capacity in Germany until 2050 is defined to 60 GW in 2030 and 93 GW in 2050, based on the Renewable Energy Sources Act 2017. The draft of the Renewable Energy Sources Act 2021 now already calls for a capacity of 71 GW onshore wind in 2030[23]. This study therefore assumes about 105 GW onshore wind capacity in 2050. For the planning of the future onshore wind locations a brownfield approach is chosen, assuming that the locations and capacities of today's existing onshore wind turbines will not change. This means that with a current capacity of about 55 GW[23], 50 GW would have to be added by 2050. The methodology of this study explores what an expansion might look like under different possible objectives. It draws on a range of publicly available datasets to ensure reproducibility of the approach. First, we describe the datasets used, before we present the multi-criteria planning approach developed.

**Open data**

This study makes use of several mainly publicly available data sets. The data sets on the future onshore wind turbine potential (1), on the existing turbines and transformers (2) and on the scenicness quality values (3) are described below.

**1) Wind turbine potential**

In Ryberg et al.[38], the future onshore wind energy potential throughout Europe was determined, with the turbines being placed at exact locations throughout Europe. According to Ryberg et al.[38], approximately 160,000 turbines with a capacity of 620 GW and an annual energy yield of 1,330 TWh can be placed in Germany. The results of the study are freely available[70] and are suitable for the present study, as future-oriented assumptions on turbine cost and design for 2050 are made. The data includes capacity, full-load hours and LCOEs for each turbine. In the present study, the mean full load hours from all considered weather years are used.

**2) Existing turbines and transformers**

For determining the locations of existing wind turbines and transformers in Germany, OpenStreetMap data are used. The existing wind turbines are identified via the Overpass API and the following query:

```
[timeout:900];
area["ISO3166-1"="DE"]->.a;
(
node["power"="generator"]["generator:source"="wind"](area.a);
way["power"="generator"]["generator:source"="wind"](area.a);
relation["power"="generator"]["generator:source"="wind"](area.a);
);
out qt;>;out qt;
```

The result can then be further processed as a geojson file in a GIS programme. In OpenStreetMap, 28,477 wind turbines are recorded (Figure 8), which corresponds to 97% of the real stock number of about 29,500 turbines[71]. These existing wind turbines are used to exclude turbine locations from the onshore wind potential of the green-field study by Ryberg et al.[38]. As in Ryberg et al.[38], regardless of the rotor diameter (which is unfortunately unknown), minimum distance buffers are drawn around the existing wind turbines, with a diameter of 1,088 m. All wind turbines from the study by Ryberg et al.[38] which are located within the ellipses of existing turbines are excluded from this analysis. This leads to the exclusion of 13.9% of turbines, 14.4% of capacity and 15.7% of energy yield of the onshore wind potential in Ryberg et al.[38].

The connection of wind turbines in Germany is mainly in the medium and high voltage levels with 96% of all turbines[72]. In the power plant database of the Federal Network Agency[73], the voltage levels of the connections are specified. Supplementary Figure S12 presents a violin plot showing which wind plants are connected to which voltage level depending on the nominal power. Most wind plants are connected to 20 kV (40%, capacity between 2.0 MW and 32.0 MW) and 110 kV (45%, capacity between 2.3 MW and 119.6 MW). Therefore, only the transformers of these two voltage levels are obtained via the Overpass API, in this example for 110 kV:

```
[timeout:900];
area["ISO3166-1"="DE"]->.a;
(
relation["power"="substation"]["voltage"~".*110000.*"](area.a);
way["power"="substation"]["voltage"~".*110000.*"](area.a);
relation["power"="sub_station"]["voltage"~".*110000.*"](area.a);
way["power"="sub_station"]["voltage"~".*110000.*"](area.a);
relation["power"="station"]["voltage"~".*110000.*"](area.a);
way["power"="station"]["voltage"~".*110000.*"](area.a);
);
out qt;>;out qt;
```

The length of the power lines connecting each wind turbine to the nearest transformer was determined using the straight line.

### 3) Scenicness data

In Roth et al.[41], for the first time scenicness quality values, i.e. values for assessing the beauty of landscapes, were determined for the whole territory of a country (Germany). The applied statistical model was based on roughly 45,000 photo-assessments of 3,500 participants, which were representatively distributed over Germany. The model explains about 64% of the variance of perceived scenicness by objectively measurable parameters and indicators such as landscape elements and land uses. To have German-wide input data for the model, standardized geo-data was used.

The scenicness values range from 1 (low scenicness) to 9 (high scenicness) and are distributed heterogeneously across the German territory (Figure 8). The highest scenicness in Germany is found in areas with steep terrain, natural landscapes and low presence of human interference[41]. These areas include the Black Forest in the southwest, the Bavarian Forest in the southeast and the Alps in the south. Areas with high human interference such as cities, on the other hand, show low scenicness. The scenicness data is the only data set not (yet) publicly available that is used in this study. However, as in the present study, the data set can be provided for scientific studies by the German Federal Agency for Nature Conservation.

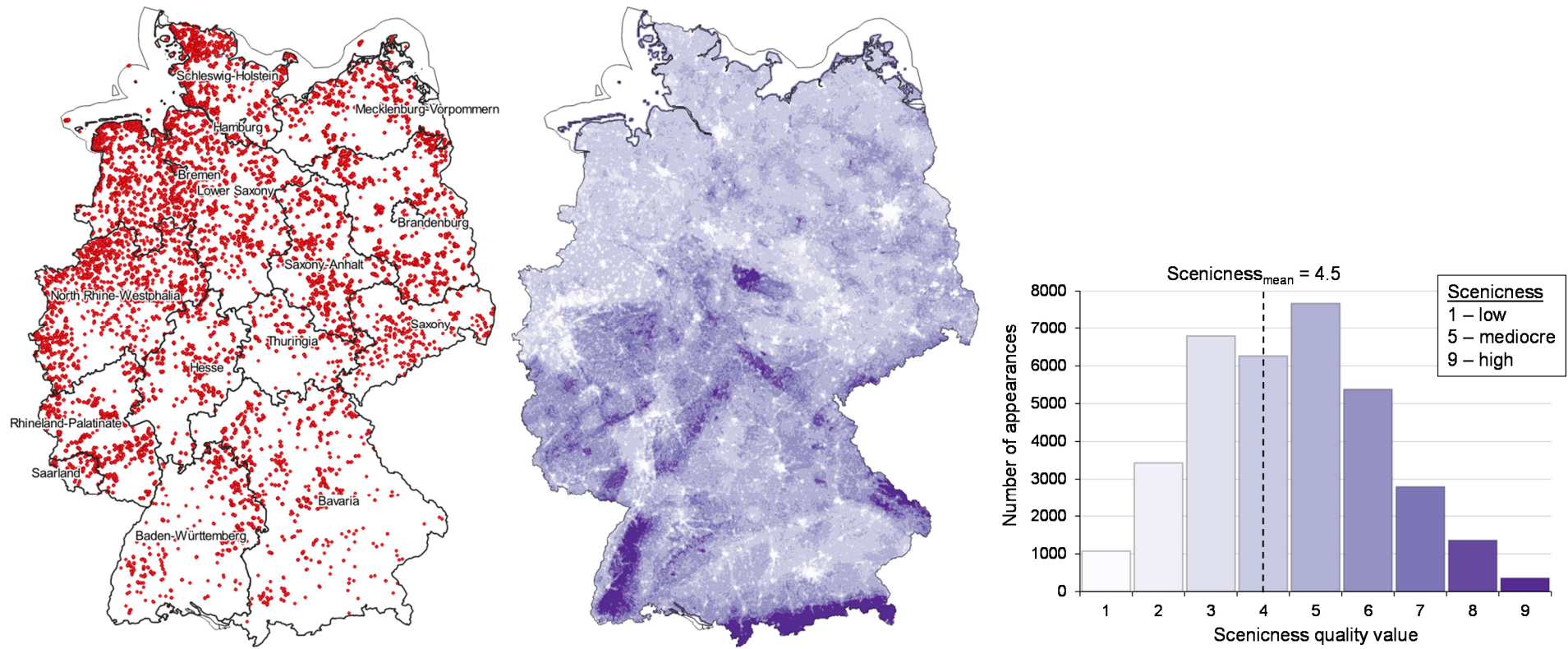

*Figure 8: Existing wind turbines in Germany and scenicness values distributed among the German territory. In the left map of Germany, the existing turbines (about 28,500) are shown as red dots. The color of the scenicness values on the right map of Germany correspond to the colours in the histogram showing the distribution and mean of these values.*

**Multi-objective optimization**

In the multi-objective optimization model developed in this study, the optimal turbine locations are selected depending on various target criteria, based on the technically feasible turbine locations for 2050 identified in Ryberg et al.[38]. Three different target criteria are taken into account in the optimization in order to show different possible expansion strategies:

1) *cost-efficiency*, as it is usually the main objective in energy system plannings[9],
2) *public acceptance*, as the plans of the energy transition are affected by a growing number of conflicts around onshore wind[12],
3) *distance to nearest transformer*, as the length of the electricity network is associated with high additional costs[35] and the necessary network could also lead through areas with high scenicness, hence affecting the public acceptance of the wind onshore project[12].

The above mentioned criteria are highly relevant for planners of national energy systems. However, a centralized expansion of onshore wind would overlook the relative costs and opportunities, which could emerge for local regions by installing and operating wind turbines[9]. Therefore, the analysis is additionally conducted for scenarios with an equitable onshore wind expansion, in which the wind turbines are distributed as evenly as possible among the different municipalities.

In the analysis we investigate scenarios, in which the above-mentioned target criteria are considered. As described above, based on the current targets of the German government, an expansion of onshore wind capacity of 50 GW for 2050 is planned in this study. However, the German Wind Association calls for an annual addition of at least 4.7 GW to meet climate targets[36]. Extrapolating this figure to 2050 results in a capacity of about 200 GW, i.e., 145 GW would have to be added. Therefore, besides the scenario with an addition of 50 GW ("Base" scenario), we also investigate a scenario with an expansion of 145 GW ("High" scenario) to assess whether an early increase in policy targets could have a large impact on the planning flexibility.

The following sections first discuss the general multi-objective optimization problem (1), then the extension for equitable expansion (2) and finally the methodology for considering all objective criteria in a single optimization (3).

**1) General model**

The parameters and variables of the multi-objective optimization model are shown in Table 2. As can be seen, the problem contains only one kind of variable, a binary variable ($b_{inst,i,j}$) for selecting the different possible turbine locations (*i*) in the various municipalities (*j*).

Table 2: Variables and parameters of the multi-objective optimization model, as well as their definitions.

| Variable / parameter | Description |
|---|---|
| $b_{inst,i,j}$ | Decides if wind turbine *i* is installed for municipality *j* (binary variable) |
| $C_{i,j}$ | LCOE of wind turbine *i* in municipality *j* (parameter) |
| $Cap_{i,j}$ | Capacity of wind turbine *i* in municipality *j* (parameter) |
| $Cap_{obj}$ | Onshore wind capacity target by 2050 (parameter) |
| $Cap_{obj,j}$ | Onshore wind capacity target in municipality j by 2050 (parameter) |
| $L_{i,j}$ | Length of electricity network to connect wind turbine *i* in municipality *j* with nearest transformer (parameter) |
| $M$ | Total number of municipalities (parameter) |
| $M_c$ | Maximum allowed total LCOE (parameter) |
| $M_l$ | Maximum allowed total length of electricity network (parameter) |
| $M_s$ | Maximum allowed total scenicness (parameter) |
| $N$ | Total number of possible turbine instalments (parameter) |
| $S_{i,j}$ | Scenicness at location of wind turbine *i* (parameter) |
| $w_c$ | Weighting factor for LCOE objective (parameter) |
| $w_l$ | Weighting factor for length of electricity network objective (parameter) |
| $w_s$ | Weighting factor for scenicness objective (parameter) |

$$\min z = \sum_{i=1}^{N}\sum_{j=1}^{M} b_{inst,i,j} \cdot (w_c \cdot C_{i,j} + w_s \cdot S_{i,j} + w_l \cdot L_{i,j}) \qquad 1$$

subject to

$$Cap_{obj} \leq \sum_{i=1}^{N}\sum_{j=1}^{M} b_{inst,i,j} \cdot Cap_{i,j} \qquad 2$$

$$\sum_{i=1}^{N}\sum_{j=1}^{M} b_{inst,i,j} \cdot C_{i,j} \leq M_c \qquad 3$$

$$\sum_{i=1}^{N}\sum_{j=1}^{M} b_{inst,i,j} \cdot S_{i,j} \leq M_s \qquad 4$$

$$\sum_{i=1}^{N}\sum_{j=1}^{M} b_{inst,i,j} \cdot L_{i,j} \leq M_l \qquad 5$$

$$b_{inst,i,j} \in \{0,1\} \qquad 6$$

In the objective function (Eq. 1), the LCOEs ($C_{i,j}$), the scenicness ($S_{i,j}$) and/or the length of the electricity network ($L_{i,j}$) are minimized. Different weights between 0 and 1 can be assigned to the target criteria using $w_c$, $w_s$ and $w_l$ to determine the relative importance of the criteria. In the analysis conducted in this article, only the values 0 or 1 are used. For an assessment of which target criterion should be given a higher/lower weight compared to the others, expert elicitations and multi-criteria decision analyses would be helpful for future analyses (see discussion section).

Eq. 2 ensures that the targeted capacity expansion ($Cap_{obj}$) is achieved by installing turbines in the various German municipalities. In Eqs. 3-5, the maximum permitted total values for LCOEs ($M_c$), scenicness ($M_s$) or network length ($M_l$) can be defined. This enables the determination of pareto curves, in which the changes of the individual objective values are shown in dependence on each other.

2) **Including Equity**

Eq. 7 is introduced to take into account an equitable expansion of the turbines. The capacity $Cap_{obj,j}$ represents the onshore wind capacity that should at least be added in a municipality $j$ in order to achieve as much equity as possible. This capacity is calculated by multiplying the share of the population of a municipality in the total German population by the capacity target for Germany. The capacity of the existing turbines in a municipality is subtracted from the result. If the potential in a municipality is not sufficient to be greater than $Cap_{obj,j}$, then $Cap_{obj,j}$ is reduced to the maximum achievable value. Otherwise, the optimization problem would not be solvable.

$$Cap_{obj,j} \leq \sum_{i=1}^{N} b_{inst,i,j} \cdot Cap_{i,j} \qquad \forall j = 1, \ldots, M \qquad 7$$

As in recent studies[32–34], we use the Gini index[43] to measure regional equity, i.e. how even the wind turbines are distributed. We adopt the formulation of Sasse and Trutnevyte[34], who adapted the Gini index as follows:

$$Regional\ equity = 1 - Gini\ index = 1 - \frac{\sum_{j=1}^{M} \sum_{k=1}^{M} |x_j - x_k|}{2 \cdot M^2 \cdot \bar{x}} \qquad 8$$

In this case, 100% means the highest and 0% the lowest regional equity score. Thereby, *x* is the capacity of wind turbines per inhabitant in municipality *j* or *k*, and M represents the total number of municipalities.

3) **Optimizing all criteria**

As described above, expert assessments would be necessary to assign appropriate weights to the target criteria. Nevertheless, this study also considers a scenario in which all target criteria are considered simultaneously. For this purpose, an attempt is made to weight the target criteria equally. For this purpose, the values for LCOEs, scenicness and network length (*x*) are scaled to the value *z* on the basis of their minimum values $x_{min}$ and maximum values $x_{max}$[74]:

$$z = \frac{x - x_{min}}{x_{max} - x_{min}} \qquad 9$$

However, even with such an approach, the distributions of the individual target criteria could deviate greatly from one another, if, for example, outliers cause most of the scaled values of a target criterion to be very low and thus not to be significant in the optimization. Therefore, the distributions of all target criteria were subsequently adjusted with a scaling factor so that all distributions have the same mean value. The resulting scaled value distributions of all target criteria are shown in a histogram in Supplementary Figure S13.

**Supplementary Figures**

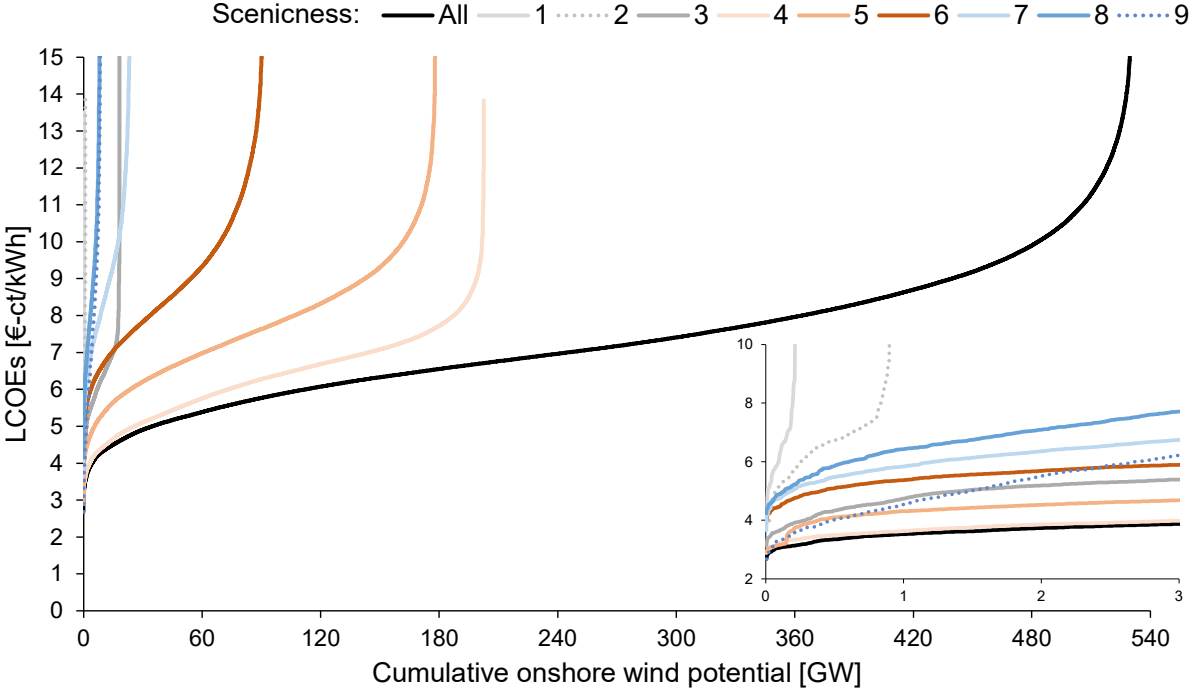

*Figure S9: Cumulative curve of the technically feasible onshore wind potential in Germany as a function of LCOEs and divided according to the scenicness of the landscape in which the corresponding wind turbines would be located. The y-axis has been clipped for better visibility of the plot at 15 €-ct/kWh. Also for better visibility, the small diagram on the right shows the curves for a low cumulative potential up to 3 GW.*

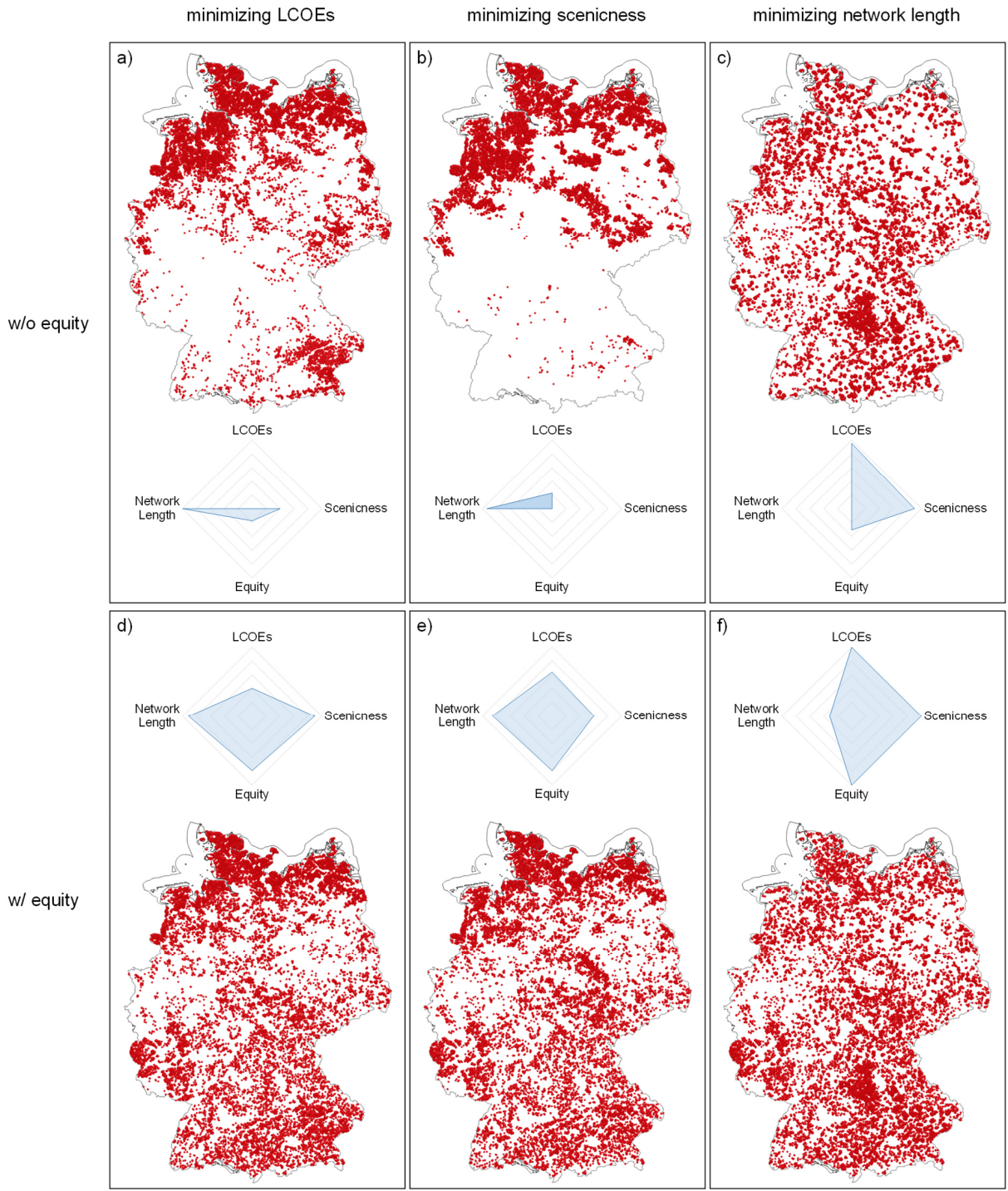

*Figure S10: Optimal locations of onshore wind turbines to be added by 2050 for different target criteria. Turbines are shown as red dots. Around 145 GW of capacity is added in each case. a) – c) show the optimal locations when minimizing LCOEs (scenario High_LCOE), scenicness (High_Scenic) or network length (High_Network), respectively. In d) – f) the same target criteria are used, but in these cases with the constraint, that the capacity expansion has to be equitable (High_LCOE_E, High_Scenic_E and High_Network_E). The values in the spider charts are scaled based on the minimum and maximum values among the "High" scenarios.*

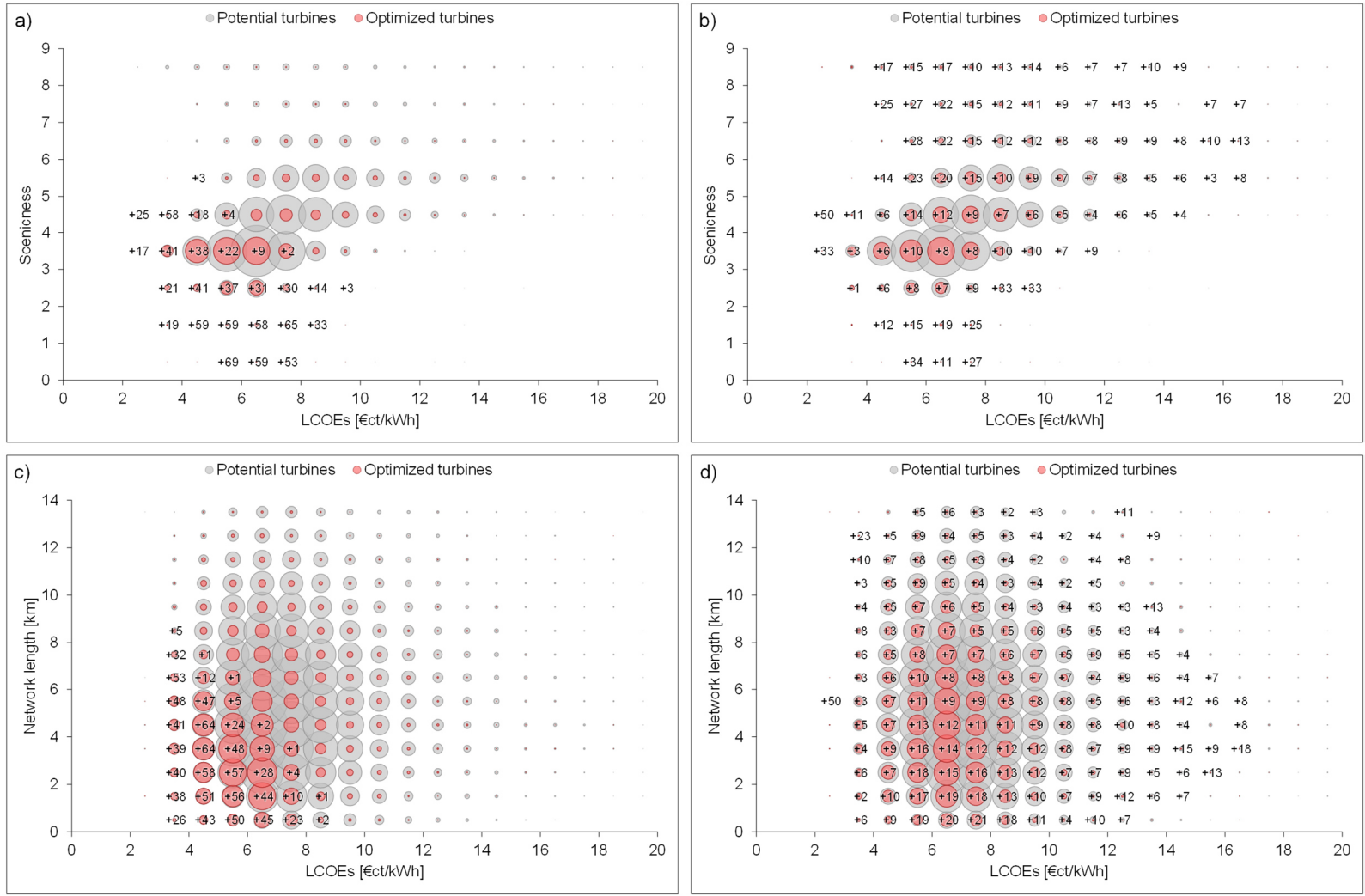

*Figure S11: Number of turbines that could potentially be placed in Germany in 2050 (gray bubbles) and the associated ranges of LCOEs (a, b, c and d), scenicness (a and b) and network length (c and d). The red bubbles show the number of turbines installed in 2050 in scenarios Base_all (a and c) and Base_all_E (b and d) and the numbers in the center of the bubbles show the increase in their share in the total potential compared to the turbine stock in 2019, i.e. +50 means that the potential has been increased by 50%. Each bubble applies to an interval, i.e. a bubble between the values 1 and 2 shows the number of turbines at the range (1;2].*

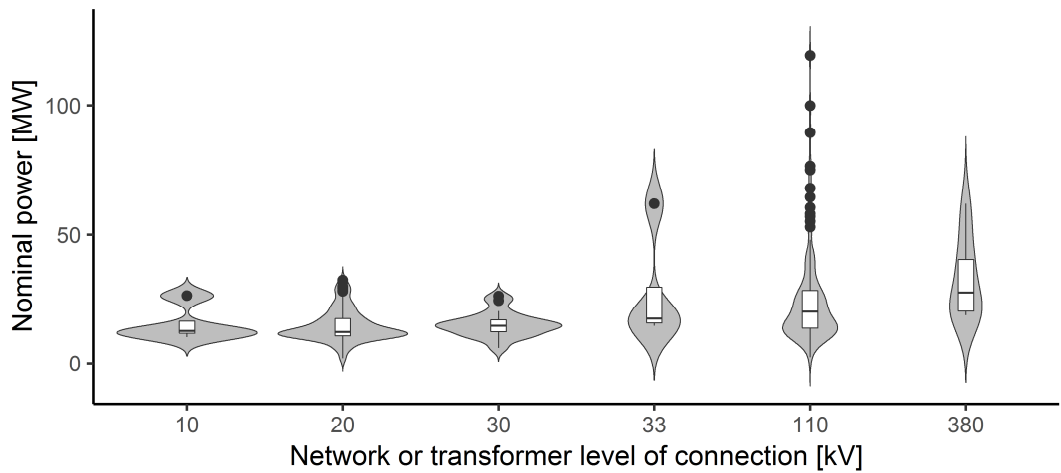

*Figure S12: Violin plot of the network or transformer level of connection for existing wind turbines, depending on their nominal power. The plot is based on data from the Federal Network Agency[73]. The width of the grey areas indicates the frequency of a nominal power. Box plots are also included in the centre of the areas, in order to show median, lower and upper quartiles as well as outliers.*

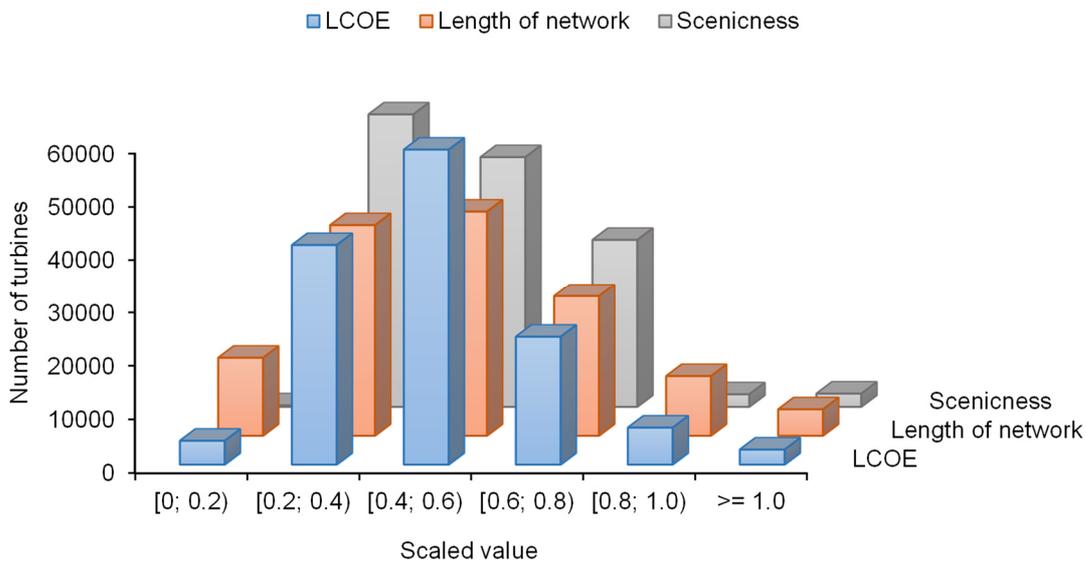

*Figure S13: Histogram of the scaled values of the target criteria LCOE, scenicness and network length. These values are used in the optimization, which incorporates all target criteria, equally weighted.*

**Acknowledgement**

**Author contributions**

Conceptualization, J.W., R.M.; Methodology, J.W., R.M.; Formal Analysis, J.W.; Data Curation, J.W.; Writing – Original Draft, J.W., R.M., H.H., and M.R.; Writing – Review and Editing, R.M., J.W.; Writing – Interactive Feedback, R.M., H.H., M.R., D.S., and W.F.; Visualization, J.W..

18. Wolsink, M. Co-production in distributed generation: renewable energy and creating space for fitting infrastructure within landscapes. *Landscape Research* **43,** 542–561; 10.1080/01426397.2017.1358360 (2018).

19. Molnarova, K. *et al.* Visual preferences for wind turbines: Location, numbers and respondent characteristics. *Applied Energy* **92,** 269–278; 10.1016/j.apenergy.2011.11.001 (2012).

20. Spielhofer, R., Hunziker, M., Kienast, F., Wissen Hayek, U. & Grêt-Regamey, A. Does rated visual landscape quality match visual features? An analysis for renewable energy landscapes. *Landscape and Urban Planning* **209,** 104000; 10.1016/j.landurbplan.2020.104000 (2021).

21. IEA. Renewables 2020. Analysis and forecast to 2025. Wind. Available at https://www.iea.org/reports/renewables-2020/wind (2021).

22. Statista. Global onshore wind energy capacity in 2019, by country. Available at https://www.statista.com/statistics/476318/global-capacity-of-onshore-wind-energy-in-select-countries/ (2020).

23. Clean Energy Wire. Ministry plans renewables expansion push to reach Germany's 2030 target. Available at https://www.cleanenergywire.org/news/ministry-plans-renewables-expansion-push-reach-germanys-2030-target (2020).

24. Statista. Approximate wind energy penetration in leading wind markets in 2019, by select country. Available at https://www.statista.com/statistics/217804/wind-energy-penetration-by-country/ (2020).

25. Clean Energy Wire. German onshore wind power – output, business and perspectives. Available at https://www.cleanenergywire.org/factsheets/german-onshore-wind-power-output-business-and-perspectives (2020).

26. Clean Energy Wire. Limits to growth: Resistance against wind power in Germany. Available at https://www.cleanenergywire.org/factsheets/fighting-windmills-when-growth-hits-resistance (2019).

27. Buck, T. Germans fall out of love with wind power. *Financial Times* (2019).

28. Weinand, J. M., McKenna, R., Kleinebrahm, M., Scheller, F. & Fichtner, W. The Impact of Public Acceptance on Cost-Efficiency and Environmental Sustainability in Decentralized Energy Systems. *Patterns. In press* (2021).

29. Lehmann, P. *et al.* Managing spatial sustainability trade-offs: The case of wind power. *Ecological Economics* **185,** 107029; 10.1016/j.ecolecon.2021.107029 (2021).

30. Harper, M., Anderson, B., James, P. & Bahaj, A. Assessing socially acceptable locations for onshore wind energy using a GIS-MCDA approach. *International Journal of Low-Carbon Technologies* **14,** 160–169; 10.1093/ijlct/ctz006 (2019).

31. Lombardi, F., Pickering, B., Colombo, E. & Pfenninger, S. Policy Decision Support for Renewables Deployment through Spatially Explicit Practically Optimal Alternatives. *Joule* **4,** 2185–2207; 10.1016/j.joule.2020.08.002 (2020).

32. Drechsler, M. *et al.* Efficient and equitable spatial allocation of renewable power plants at the country scale. *Nat Energy* **2**; 10.1038/nenergy.2017.124 (2017).

33. Sasse, J.-P. & Trutnevyte, E. Distributional trade-offs between regionally equitable and cost-efficient allocation of renewable electricity generation. *Applied Energy* **254,** 113724; 10.1016/j.apenergy.2019.113724 (2019).